# Group-III doping of the p-type oxide SnO

View Online Export Citation CrossMark

N. Gutmann,[a] G. Hoffmann, A. F. Campbell, J. Lähnemann, and O. Bierwagen[a]

## AFFILIATIONS

Paul-Drude-Institut für Festkörperelektronik, Leibniz-Institut im Forschungsverbund Berlin e.V., Hauvogteiplatz 5-7, 10117 Berlin, Germany

[a]Authors to whom correspondence should be addressed: gutmann@pdi-berlin.de and bierwagen@pdi-berlin.de

## ABSTRACT

Al-, Ga-, and In-doped tin monoxide (SnO) films were grown by molecular beam epitaxy. Hall measurements reveal acceptor character of all investigated group-III dopants in SnO, which corresponds to theoretical calculations and previous experiments for Ga and In, but contradicts theoretical calculations predicting donor character for Al in SnO. Doping concentrations of $10^{18}$–$10^{22}$ cm$^{-3}$ were tested and maximum hole concentrations of $2 \times 10^{19}$, $3 \times 10^{19}$, and $1 \times 10^{19}$ cm$^{-3}$ with doping efficiencies of 3%, 9%, and 1% were found for Al, Ga, and In, respectively. The observed doping limit in the mid $10^{19}$ cm$^{-3}$ aligns with the limits observed in literature for different dopants, hinting to fundamental aspects of the SnO defect chemistry. Due to the unexpected Al acceptor behavior, we hypothesized $Al^{1+}$ being present instead of $Al^{3+}$. To stabilize the higher $Al^{3+}$ oxidation state, another sample series of Al-doped SnO was grown in an oxidizing atmosphere. Despite the oxidizing atmosphere, these samples showed p-type character. A more complex defect chemistry is thus expected to explain the Al acceptor character.

Over the last few years, transparent semiconducting oxides (TSOs) have emerged as one of the fundamental building blocks for new device concepts and serve as a foundation for novel functional physical properties.[1–4] More recently, TSOs are considered a promising material class for advanced memory architectures, in which they form an additional back-end-of-line (BEOL) compatible logic and memory layer.[5,6] Until now, the use of TSOs has been restricted to devices using n-type doped oxides, inhibiting further development of transparent oxide-based homo-pn-junctions and CMOS technology. P-type conductivity remains a major bottleneck for TSOs, as the localized O 2p-derived valence bands, represented by low in energy and shallow valence-band maxima, lead to self-compensation and low possible hole mobilities. In contrast, the valence band of the $\alpha$-phase of tin(II)monoxide ($\alpha$-SnO), first grown by Pan and Fu,[7] mainly consists of Sn 5s orbitals,[8] which are more disperse and higher in energy than O 2p orbitals. Hole mobilities ranging from 2 to 20 cm$^2$/V s measured by the Hall method have been reported in the literature,[6,9–12] and p-type-transistors have been realized.[6,10,13] Its low synthesis temperature makes SnO a suitable p-type TSO for BEOL application.[5,9] The low electronic indirect bandgap of 0.6 eV[4,14] does not inhibit transparency as this transition is optically forbidden, leading to a higher optical bandgap reported between 2.6 and 3.2 eV.[4] SnO shows thus the unique combination of a small electronic bandgap, opening the possibility for ambipolar doping[15–19] with a high optical bandgap, useful for transparent electronics. Undoped SnO naturally shows p-type behavior.[9,14,20] While the underlying mechanisms of intrinsic p-type conductivity are still under debate, density functional theory (DFT) calculations suggest the formation of Sn vacancies ($V_{Sn}$),[21,22] including their complex with hydrogen impurities.[16]

Inspired by the possibility of p-doping SnO using Ga,[23,24] Na,[24] K,[25,26] or Y,[27] the realization of an all-SnO-based pn-junction emerged as central topic. However, the search of a suitable n-type dopant turned out to be a challenging task with contradicting results. While Hosono *et al.* report successful n-doping by Sb,[18] Guo *et al.* report p-dopant behavior for Sb in SnO.[28] La was found to act as compensating donor dopant in SnO,[23] making the material more insulating but not n-type conductive. As another promising candidate, Al has been theoretically predicted to show donor behavior,[19,23] but no experimental Al-doping studies have been performed on phase-pure SnO yet. The Al-doping studies carried out on phase mixtures of SnO and the n-type conducting $SnO_2$ show conflicting data. Some groups report higher conductivity or mobility upon Al-doping,[29,30] and another group reports compensating donor behavior.[31] Table I summarizes the above-mentioned experimental SnO doping studies.

In this work, we comparatively explore doping of molecular beam epitaxy (MBE)-grown phase-pure SnO by the group-III elements Al, Ga, and In. Ga and In are both predicted by *ab initio* calculations to act as acceptors.[19,23] The acceptor character of Ga has additionally been shown experimentally,[23] while no experimental data on In-doping has been published yet. As Al is predicted to act as donor,[19] it is of special interest for ambipolar doping of SnO.

**TABLE I.** Previously published doping studies of SnO. Used doping element, unintentionally doped (UID) hole charge carrier density, maximum hole charge carrier density reached by doping, corresponding doping concentration, hole mobility, growth technique, and crystallinity and phase composition of the samples. While most studies report the Hall mobility, the two Al-doping studies report the field-effect mobility, marked by (FE) in the table.

| References [First author (year)] | Doping element | UID hole density ($cm^{-3}$) | Maximum hole density ($cm^{-3}$) | Corresponding doping concentration | Mobility ($cm^2/V\,s$) | Growth technique | Crystallinity | Phase composition |
|---|---|---|---|---|---|---|---|---|
| Guo *et al.*[28] | Y | $1.1\times10^{18}$ | $2.7\times10^{18}$ | 5 at. % | 0.6–0.9 | EBE | Poly-crystalline | SnO |
| | Sb | $1.1\times10^{18}$ | $1.3\times10^{18}$ | 5 at. % | 0.6–0.9 | EBE | Poly-crystalline | SnO |
| Hosono *et al.*[18] | Sb | $2.5\times10^{17}$ | n-type | 10 at. % | 2.3 | PLD | Single-crystalline | SnO |
| Liang *et al.*[27] | Y | $5.6\times10^{16}$ | $4.7\times10^{16}$ | 0.8 at. % | 0.8–3.9 | EBE | Poly-crystalline | SnO |
| Kwok *et al.*[24] | Ga | $4.6\times10^{18}$ | $6.2\times10^{18}$ | 1.20% | 3.6 | Sputtering | Poly-crystalline | SnO |
| | Na | $4.6\times10^{18}$ | $4.7\times10^{19}$ | 2.80% | 11.3 | Sputtering | Poly-crystalline | SnO |
| Egbo *et al.*[23] | Ga | $6\times10^{18}$–$1\times10^{19}$ | $5\times10^{19}$ | $4\times10^{20}$ $cm^{-3}$ | 4–6 | MBE | Single-crystalline | SnO |
| | La | $6\times10^{18}-1\times10^{19}$ | … | $1\times10^{20}$ $cm^{-3}$ | … | MBE | Single-crystalline | SnO |
| Chae *et al.*[25] | K | $4.8\times10^{17}$ | $1.5\times10^{19}$ | $<5.3\times10^{19}$ $cm^{-3}$ | 3.7 | MBE | Single-crystalline | SnO |
| Nguyen *et al.*[29] | Al | Not provided | Not provided | 1.10% | 1.85 (FE) | Sputtering | Poly-crystalline | $SnO + SnO_2$ |
| Pan *et al.*[30] | Al | Not provided | Not provided | 1% | 1.17 (FE) | ALD | Poly-crystalline | $SnO + SnO_2$ |
| Peek *et al.*[31] | Al | Not provided | Not provided | Not provided | Not provided | ALD | Poly-crystalline | $SnO + SnO_2$ |

Unintentionally doped (UID), Al-, Ga-, and In-doped, 30–55 nm-thick SnO films were grown on $5 \times 5 \times 0.5\,mm^3$ single-side polished, yttrium-stabilized zirconia (YSZ) (001) substrates, known to enable epitaxy of single-crystalline SnO(001),[20] by suboxide molecular beam epitaxy (S-MBE). To prepare well-defined, epi-ready substrate surfaces, the YSZ substrates were wiped with a cotton swab in an isopropanol bath and subsequently baked out for 1 h at 1050 °C in a tube furnace, supplying 150 sccm $O_2$ flux. As a consequence of the annealing procedure, all substrates showed a step-terrace surface, as shown in Fig. S18 of the supplementary material. The step height of around 0.5 nm is in accordance with the lattice constant of the YSZ substrate of 0.51 nm, indicating singly terminated and epi-ready surfaces. All substrates received an $SrRuO_3$ backside coating prior to growth for homogeneous heat absorption from the substrate heater. For the S-MBE growth process, an SnO suboxide source containing an $Al_2O_3$ crucible filled with a mixture of $SnO_2$ and Sn was used.[9,32] The primary flux species for the ($SnO_2 +$ Sn) cell is SnO and its oligomers, ensuring a perfect stoichiometry of the grown films. The ($SnO_2 +$ Sn) cell temperature was kept at 850 °C and the hot-lip at 950 °C, leading to beam equivalent pressures (BEPs) of $(0.8 - 1.4) \times 10^{-7}$ mbar, which correspond to growth rates in the range of 0.5–0.8 nm/min. The doping of the SnO layers was achieved by the co-deposition of the respective dopant element evaporated from an elemental source material. The necessary dopant cell temperatures to achieve a nominal dopant concentration in the film between $1 \times 10^{18}$ and $1 \times 10^{22}\,cm^{-3}$ were extrapolated from temperature-dependent BEP measurements at higher cell temperatures, as explained in the supplementary material. The corresponding cell temperatures for the Al, Ga, and In cells were 700–1000, 530–620, and 475–575 °C, respectively. The substrate temperature ($T_{sub}$) was measured by a thermocouple located behind the substrate heater and kept at 375 °C for the Al- and the In-doped samples and at 400 °C for the Ga-doped samples. In the earlier Ga-doping study published by our group,[23] the doping series grown by the plasma-assisted MBE was grown at $T_{sub} = 350$ °C, while the doping series grown by the suboxide MBE was grown at $T_{sub} = 400$ °C. As no substantial difference in the doping behavior of Ga is observed between these two series, we believe the Ga-doping at $T_{sub} = 400$ °C in the present study to adequately represent the Ga-doping behavior at $T_{sub} = 375$ °C. During SnO deposition, the growth chamber pressure was below $5 \times 10^{-9}$ mbar, confirming the absence of parasitic oxygen emission from the suboxide source. The growth was *in situ* monitored by reflection high-energy electron diffraction (RHEED) (see Fig. S1 of the supplementary material).

For electrical characterization, the films were contacted by In-coated Cu-probes and measured in van der Pauw geometry.[33] A magnetic field of ±800 mT and currents in the range of 50–100 $\mu$A were used. For the analysis of the crystal structure, a four-circle x-ray lab-diffractometer (X'pert Pro MRD, Philips PANalytical) using Cu K–$\alpha$ radiation was used. Symmetric out-of-plane $2\theta$–$\omega$ scans and rocking curves of the SnO 002 reflection were taken using a 1/2° entrance slit and a 1 mm detector slit. Additionally, high-resolution scans of the SnO 002 reflection were performed using an analyzer crystal. The SnO layer thicknesses were determined from thickness fringes observed in high-resolution x-ray diffraction (HR-XRD) or by x-ray reflectivity (XRR) if no fringes were observed, as shown in Figs. S9 and S13–S17 of the supplementary material, respectively. The surface morphology was determined by atomic force microscopy (AFM) in peak force tapping mode using a Bruker Dimension Edge. Energy-dispersive x-ray spectroscopy (EDX) measurements were performed in a scanning electron microscope using an EDAX Octane Elect Super detector on selected samples with the highest doping concentrations. To account for contributions from both the thin film and substrate material in the depth probed, the dopant concentrations were obtained from measurements at different acceleration voltages and analyzed using a standards-based quantification routine implemented in the software BadgerFilm.[34] Spectra were acquired over $13 \times 16\,\mu m^2$ areas to average out any lateral variations. Further details on this quantification routine and related challenges can be found in section IV of the supplementary material. Table II provides an overview of the grown and investigated samples along with growth parameters and characterization results.

**TABLE II.** Overview of growth parameters and characterization results of all unintentionally doped (UID), Al-, Ga, and In-doped samples discussed in this work: Doping cell temperature $T_{cell}$, nominal doping $N_D$, substrate temperature $T_{sub}$, beam equivalent pressure of the ($SnO_2$ + Sn)-cell BEP, layer thickness $t$, Hall resistance $\rho_{Hall}$, hole mobility $\mu$, resistance $\rho$, measured hole density $N_h$, full width at half maximum of the XRD rocking curve and the 2Θ–ω scan of the SnO 002 reflection $\Delta\omega$ and $\Delta 2\Theta$, respectively, and the root mean square (rms) roughness of the AFM scans $r$. $d$ denominates the error of the respective value. The sample thickness $t$ is determined from the thickness fringes of the high-resolution XRD or XRR scans (marked with *), found in Figs. S9 and S13–S17, respectively, of the supplementary material. The hole density is calculated from the Hall resistance $\rho_{Hall}$ and the sample thickness $t$.

| Doping element | $T_{cell}$ (°C) | $N_D$ ($cm^{-3}$) | $T_{sub}$ (°C) | SnO BEP ($10^{-7}$ mbar) | $t$ (nm) | $d_t$ (nm) | $\rho_{Hall}$ (Ω/T) | $d_{\rho_{Hall}}$ (Ω/T) | $\mu$ ($\frac{cm^2}{Vs}$) | $\rho$ (Ω cm) | $d_\rho$ (Ω cm) | $N_h$ ($cm^{-3}$) | $d_{N_h}$ ($cm^{-3}$) | $\Delta\omega$ (°) | $\Delta 2\Theta$ (°) | $r$ (nm) |
|---|---|---|---|---|---|---|---|---|---|---|---|---|---|---|---|---|
| UID | … | 0 | 400 | 0.9 | 30 | 5 | +68.3 | 14.7 | 1.9 | 1.10 | 0.18 | $3.1\times10^{18}$ | $8\times10^{17}$ | 0.14 | 0.309 | 12.4 |
| | … | 0 | 375 | 1.3 | 45 | 5 | +75.2 | 7.5 | 3.3 | 1.03 | 0.11 | $1.9\times10^{18}$ | $3\times10^{17}$ | 0.12 | 0.273 | 19.9 |
| Al | 700 | $1.4\times10^{18}$ | 375 | 1.2 | 55 | 10 | +80.0 | 13.0 | 4.6 | 0.96 | 0.17 | $1.4\times10^{18}$ | $3\times10^{17}$ | 0.21 | 0.241 | … |
| | 750 | $8.7\times10^{18}$ | 375 | 1.2 | 42* | 10 | +30.4 | 6.4 | 4.8 | 0.35 | 0.06 | $4.9\times10^{18}$ | $2\times10^{18}$ | 0.39 | 0.284 | 8.8 |
| | 800 | $4.7\times10^{19}$ | 375 | 1.2 | 50 | 10 | +13.9 | 1.4 | 4.1 | 0.17 | 0.03 | $9.0\times10^{18}$ | $2\times10^{18}$ | 0.19 | 0.289 | 14.4 |
| | 850 | $2.2\times10^{20}$ | 375 | 1.2 | 51 | 10 | +6.0 | 0.2 | 4.1 | 0.08 | 0.01 | $2.0\times10^{19}$ | $4\times10^{18}$ | 0.29 | 0.242 | 2.0 |
| | 900 | $8.8\times10^{20}$ | 375 | 1.3 | 56 | 5 | +4.6 | 0.3 | 1.9 | 0.13 | 0.01 | $2.4\times10^{19}$ | $3\times10^{18}$ | 0.17 | 0.221 | 7.6 |
| | 925 | $1.7\times10^{21}$ | 375 | 1.1 | 51* | 10 | +8.5 | 0.7 | 1.5 | 0.32 | 0.06 | $1.3\times10^{19}$ | $3\times10^{18}$ | 0.59 | 0.297 | 0.8 |
| | 950 | $3.2\times10^{21}$ | 375 | 1.1 | 57* | 10 | +14.0 | 3.9 | 1.4 | 0.57 | 0.10 | $7.8\times10^{18}$ | $3\times10^{18}$ | 0.83 | 0.328 | … |
| | 1000 | $1.0\times10^{22}$ | 375 | 1.4 | 55 | 10 | … | … | … | 53.55 | 10.71 | … | … | … | 0.555 | … |
| Ga | 530 | $3.3\times10^{18}$ | 400 | 0.9 | 30 | 5 | +34.6 | 6.3 | 3.1 | 0.33 | 0.06 | $6.0\times10^{18}$ | $1\times10^{18}$ | 0.14 | 0.326 | … |
| | 560 | $1.1\times10^{20}$ | 400 | 0.9 | 30 | 5 | +10.0 | 0.6 | 3.9 | 0.08 | 0.01 | $2.1\times10^{19}$ | $4\times10^{18}$ | 0.19 | 0.295 | 9.8 |
| | 590 | $3.6\times10^{20}$ | 400 | 0.9 | 30 | 5 | +7.1 | 0.2 | 3.8 | 0.06 | 0.01 | $2.9\times10^{19}$ | $5\times10^{18}$ | 0.11 | 0.240 | 3.1 |
| | 620 | $1.0\times10^{21}$ | 400 | 0.9 | 34* | 5 | +9.7 | 0.9 | 4.2 | 0.07 | 0.01 | $1.9\times10^{19}$ | $4\times10^{18}$ | 0.12 | 0.293 | … |
| In | 475 | $8.6\times10^{18}$ | 375 | 1.1 | 45 | 5 | +41.9 | 11.7 | 4.3 | 0.44 | 0.05 | $3.3\times10^{18}$ | $1\times10^{18}$ | 0.13 | 0.237 | … |
| | 525 | $5.2\times10^{20}$ | 375 | 1.1 | 47 | 5 | +13.7 | 0.6 | 2.6 | 0.25 | 0.03 | $9.7\times10^{18}$ | $1\times10^{18}$ | 0.10 | 0.201 | 5.9 |
| | 575 | $2.6\times10^{21}$ | 375 | 1.1 | 50 | 5 | +10.1 | 1.3 | 1.0 | 0.50 | 0.05 | $1.2\times10^{19}$ | $2\times10^{18}$ | 0.11 | 0.240 | … |
| Al in $O_2$ | … | 0 | 375 | 1.2 | 55 | 5 | +50.3 | 12.7 | 4.7 | 0.59 | 0.05 | $2.3\times10^{18}$ | $6\times10^{17}$ | 0.34 | 0.288 | … |
| | 750 | $8.7\times10^{18}$ | 375 | 1.1 | 37 | 5 | +41.4 | 4.0 | 4.5 | 0.34 | 0.05 | $4.1\times10^{18}$ | $7\times10^{17}$ | 0.13 | 0.246 | … |
| | 850 | $2.2\times10^{20}$ | 375 | 1.1 | 37 | 5 | +3.7 | 1.0 | 3.6 | 0.04 | 0.01 | $4.6\times10^{19}$ | $1\times10^{19}$ | 0.17 | 0.250 | 2.6 |
| | 900 | $8.8\times10^{20}$ | 375 | 1.1 | 45 | 10 | +34.6 | 1.3 | 2.5 | 0.16 | 0.04 | $1.6\times10^{19}$ | $4\times10^{18}$ | 0.19 | 0.284 | … |

All investigated samples exhibit positive Hall resistances, as shown in Table II, demonstrating the p-type character irrespective of the dopant used in this study. Figure 1(a) summarizes the measured hole density of the Al-, Ga-, and In-doped SnO as function of nominal dopant concentration. The purple horizontal lines mark the measured hole densities of the two unintentionally doped (UID) SnO reference samples, with the purple shaded area describing their error. These UID hole concentrations are higher compared to values of other groups who achieve UID hole concentrations down to $10^{17}\,cm^{-3}$[25,28] and $10^{16}\,cm^{-3}$.[14] After this study, we replaced the source material by a fresh $SnO_2 + Sn$ mixture and observed lower unintentional hole concentrations in the mid-$10^{17}\,cm^{-3}$. Consequently, we attribute the higher UID hole concentration of the present doping study to a cross contamination of the source material with acceptor dopant species. More importantly, Fig. 1(a) shows that the hole concentration of the Al-, Ga-, and In-doped films increases with the nominal dopant concentrations up to a certain maximum value shown in Table III, proving acceptor dopant type behavior for all the tested group-III dopants.

In terms of dopant type, these findings are qualitatively in line with theoretical predictions for Ga and In (i.e., being present in the $Ga^{1+}/In^{1+}$ state on the $Sn^{2+}$ site),[23] yet contradict the theoretical prediction of Al to act as a donor (i.e., being present in the $Al^{3+}$ state on the $Sn^{2+}$ site).[19,23] For Ga-doping, our results are quantitatively in line with previous experimental reports: The maximum hole concentration of $(2.9 \pm 0.5) \times 10^{19}\,cm^{-3}$ shown here aligns well with the value of $5 \times 10^{19}\,cm^{-3}$ found in the literature,[23,24] proving the reproducibility and validity of this study. Indium, showing the lowest doping efficiency, contrasts theoretical calculations, which predict In to be a more effective dopant than Ga and Al.[23] In addition to the highest measured hole concentration and the respective nominal doping concentrations for each dopant element, Table III summarizes their ratio as corresponding doping efficiency and a dopant activation energy $E_A$ derived from the charge balance equation for incomplete ionization of singly ionized acceptors,[35]

$$p \approx \sqrt{\frac{N_A N_V}{2}} \exp\left(-\frac{E_A}{2k_B T}\right), \tag{1}$$

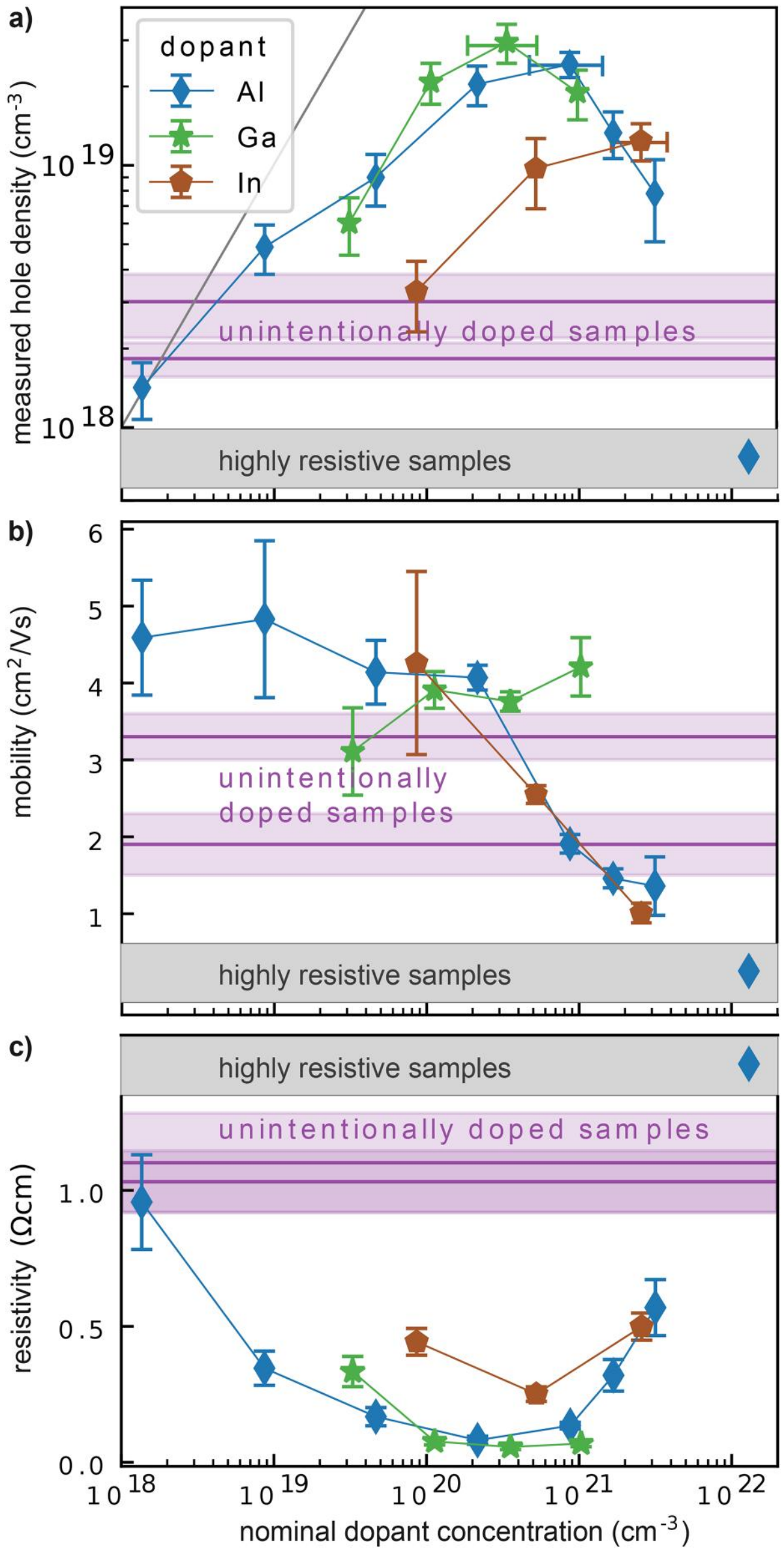

**FIG. 1.** Hole transport properties determined by Hall measurements as function of the nominal doping with Al- (blue diamonds), Ga- (green stars), In- (brown pentagons), and unintentionally doped SnO samples (purple shaded area). (a) Hole density. For comparison, the gray line denotes the nominal acceptor density. Error of the nominal doping concentration determined by the EDX shows exemplary for the highest doped samples (see supplementary material section IV, EDX). (b) Hole mobility. (c) Resistivity.

where $N_A$ denotes the acceptor concentration, $N_V$ the effective valence band density of states, $k_B$ the Boltzmann constant, and $T$ the temperature. $N_V$ is calculated from

$$N_V = 2\left(\frac{2\pi m_h^* k_B T}{h^2}\right)^{3/2}, \tag{2}$$

**TABLE III.** Highest measured hole density and respective nominal doping for Al-, Ga, and In-doped samples. Furthermore, doping efficiency and acceptor activation energy $E_A$ calculated from these values using the charge balance equation [Eq. (1)].

| Doping element | Highest measured hole density ($cm^{-3}$) | Nominal doping ($cm^{-3}$) | Doping efficiency | $E_A$ (meV) |
|---|---|---|---|---|
| Al | $(2.3 \pm 0.4) \times 10^{19}$ | $8.8 \times 10^{20}$ | 3.3% | 100 |
| Ga | $(3.0 \pm 0.5) \times 10^{19}$ | $3.5 \times 10^{20}$ | 8.5% | 60 |
| In | $(1.2 \pm 0.2) \times 10^{19}$ | $2.6 \times 10^{21}$ | 0.5% | 160 |

assuming a density-of-state hole effective mass of $m_h^* = 1.7 \cdot m_e$.[20] The low doping efficiency and corresponding high acceptor activation energy around 100 meV suggests Al, Ga, and In to be deep acceptors in SnO. This calculation of the acceptor activation energy assumes complete electrical activation of all dopant atoms. The difference between the measured hole concentration and the nominal dopant concentration may, however, also arise from incomplete dopant activation due to incorporation in electrically inactive configurations, such as interstitial sites or dopant clusters.

Beyond a nominal doping concentration of mid–high $10^{20}$ $cm^{-3}$, the measured hole density decreases for Al and Ga. Egbo *et al.* attribute the saturation of the measured hole density beyond a specific doping concentration to an incomplete ionization of the dopants due to pinning of the Fermi level at an $\varepsilon(+/0)$ charge transition level of $Ga_{Sn}$ near the valence band edge. The formation of secondary phases for even higher doping concentrations could explain the decrease in the measured hole density. The drop in measured hole density is accompanied by a drop of the mobility for Al- and In-doping and an increase in the resistivity [Figs. 1(b) and 1(c), respectively]. Notably, the electrical characterization indicates a constant mobility of $\approx$4.2 $cm^2$/V s for Al doped samples in the range of $10^{18}$–$10^{20}$ $cm^{-3}$.

The highest hole concentration reached by Ga-doping as well as K-[25] and Na-doping[24] of $(2–5) \times 10^{19}$ $cm^{-3}$ suggest a hole concentration limit of mid $10^{19}$ $cm^{-3}$ intrinsic to SnO and independent of the dopant element (compare Table I). Practically posing a limit to the conductivity of SnO, this limit could help gaining more insights into fundamental aspects of the controversially discussed defect chemistry of SnO, requiring further investigation.

As described earlier, we experimentally observed the acceptor character of Al in contradiction to its theoretically predicted donor character. Graužinytė *et al.* predicted the incorporation of $Al^{3+}$ on the $Sn^{2+}$ site.[19,23] As the "most straightforward" explanation for the observed p-type character, we considered the incorporation of $Al^{1+}$ instead of $Al^{3+}$. A more detailed reasoning for this assumption can be found in section III of the supplementary material. In an attempt to facilitate the incorporation of $Al^{3+}$, we grew another sample series of Al-doped SnO in $O_2$ environment, aiming at pre-oxidizing the elemental $Al^0$ to $Al^{3+}$. For this purpose, an $O_2$ background pressure of $2 \times 10^{-5}$ mbar was established by flowing 2 sccm of molecular oxygen into the growth chamber. These conditions should be able to oxidize all provided Al at the growth front to $Al_2O_3$, as we have shown in a separate experiment where even a lower $O_2$ flow rate of 0.3 sccm was able to oxidize Al at a growth rate of 0.6 nm/min (will be published). Lee *et al.* showed that an $O_2$-background during growth has a low influence on the hole density of undoped SnO.[11]

Figure 2(a) compares the measured hole densities for the Al-doped SnO films grown in vacuum and in oxygen. Also the Al-doped

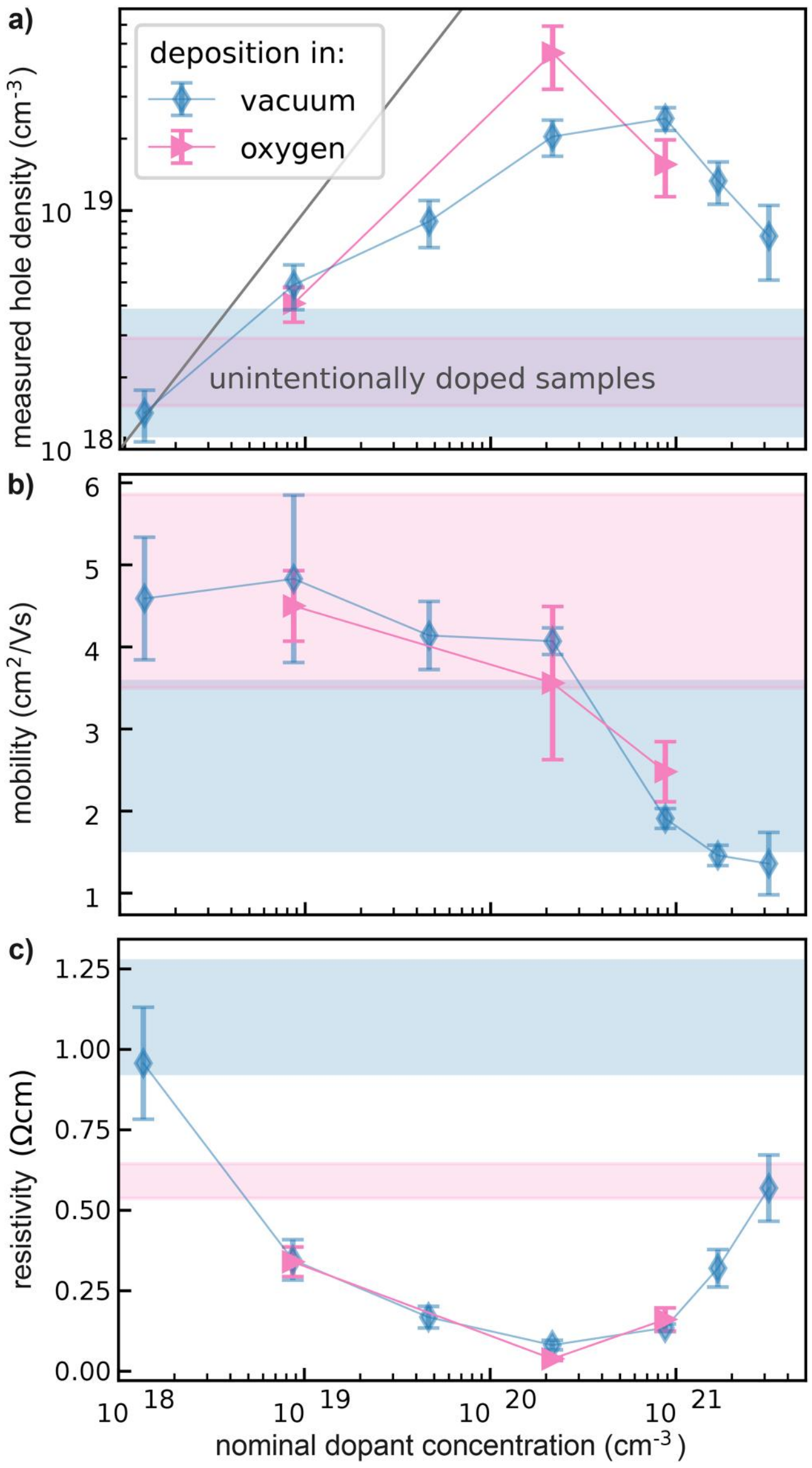

**FIG. 2.** Comparison of the electronic properties determined by Hall measurements of Al-doped SnO grown in vacuum (blue diamonds) and in $2 \times 10^{-5}$ mbar $O_2$ atmosphere (pink triangles). The blue and pink shaded areas mark the values for the unintentionally doped samples grown in vacuum and the $O_2$ atmosphere, respectively. (a) Hole density. (b) Hole mobility. (c) Resistivity.

SnO films grown in oxygen atmosphere show p-type behavior. The measured hole density over the nominal doping concentration follows the same trend as for the series grown in vacuum. Furthermore, the differences in hole mobility and resistivity between the samples grown in vacuum and oxygen atmosphere [Figs. 2(b) and 2(c)] are within the error bars. It is followed that, contrarily to theoretical predictions,[19,23] Al acts as an acceptor in SnO, even if the $Al^{3+}$ oxidation state should be supported by an oxygen growth atmosphere. This behavior calls for further study of the SnO defect chemistry beyond the scope of the present work to clarify its origin. Given the high oxidation potential of Al,[36] incomplete oxidation seems unlikely and a more complex defect reaction is assumed. For example, the complex of an $Al^{3+}$-interstitial defect with two tin vacancies would act as a singly charged acceptor $(\mathrm{Al_i^{3+}} + 2\mathrm{V_{Sn}^{2-}} \rightarrow [\mathrm{Al_i} - 2\mathrm{V_{Sn}}]^{-})$. A similar defect complex is proposed by Guo *et al.* for Sb- or Y-doping.[28]

For completeness, the structural properties of the layers are discussed: Fig. 3 shows XRD symmetric out-of-plane $2\theta$–$\omega$ scans of the undoped and doped samples. The x-ray diffractograms of all films show the characteristic SnO 001 and 002 reflections at 18.3° and 37.1°, respectively, additionally to the YSZ 002 substrate reflection at 34.9°. The absence of further reflections confirms the presence of single-phase SnO (001) layers. Only for the highest Al- and In-doping concentrations, an additional reflection at 30.7° arises, which we attribute to the Sn 200 reflection in accordance with the hypothesis of elemental Sn-clusters being present in the highly doped SnO films, as discussed in section III of the supplementary material. The SnO 002 reflection of the highest Al-doped sample shows a broad shoulder toward higher diffraction angle, indicating the presence of Al–Sn–O-related secondary phases. The full width at half maximum (FWHM) of the $\omega$ rocking curves of the SnO 002 reflection, shown in Fig. S10 of the supplementary material, are in the range of 0.1°–0.4° for Al-, Ga-, and In-doping up to $10^{21}$ $cm^{-3}$ and show no correlation with the nominal doping. These FWHM values are on the lower side of the ones reported in the literature,[11,14,20] indicating the high crystalline quality of the grown samples. For Al concentrations above $10^{21}$ $cm^{-3}$, the FWHM increases up to 0.8°, indicating structural deterioration. AFM images of the samples are shown and discussed in section VIII of the supplementary material, and their root mean square roughness is shown in Table II.

In summary, the doping of MBE-grown SnO by the group-III elements Al, Ga, and In is compared. SnO was evaporated from an $(SnO_2 + Sn)$-source, while the doping elements were evaporated from elemental sources. In-line with theoretical predictions, Ga and In are acceptors and their incorporation leads to an increased hole concentration. In contrast to its theoretically predicted donor character, Al also leads to an increased hole concentration and is thus experimentally shown to act as an acceptor. The attempt to facilitate an $Al^{3+}$ donor state by deposition in a background of molecular oxygen did not change its acceptor character, pointing toward a more complex defect chemistry of Al-doping. While dopant concentrations up to at least $10^{21}$ $cm^{-3}$ were incorporated, the corresponding hole concentrations never exceeded $3 \times 10^{19}$ $cm^{-3}$. The highest hole concentration and doping efficiency was found for Ga. The observed hole concentration limit, roughly in-line with further literature reporting maximum hole concentrations below $5 \times 10^{19}$ $cm^{-3}$,[23–25] summarized in Table I, could provide valuable information for the controversially discussed defect chemistry of SnO.

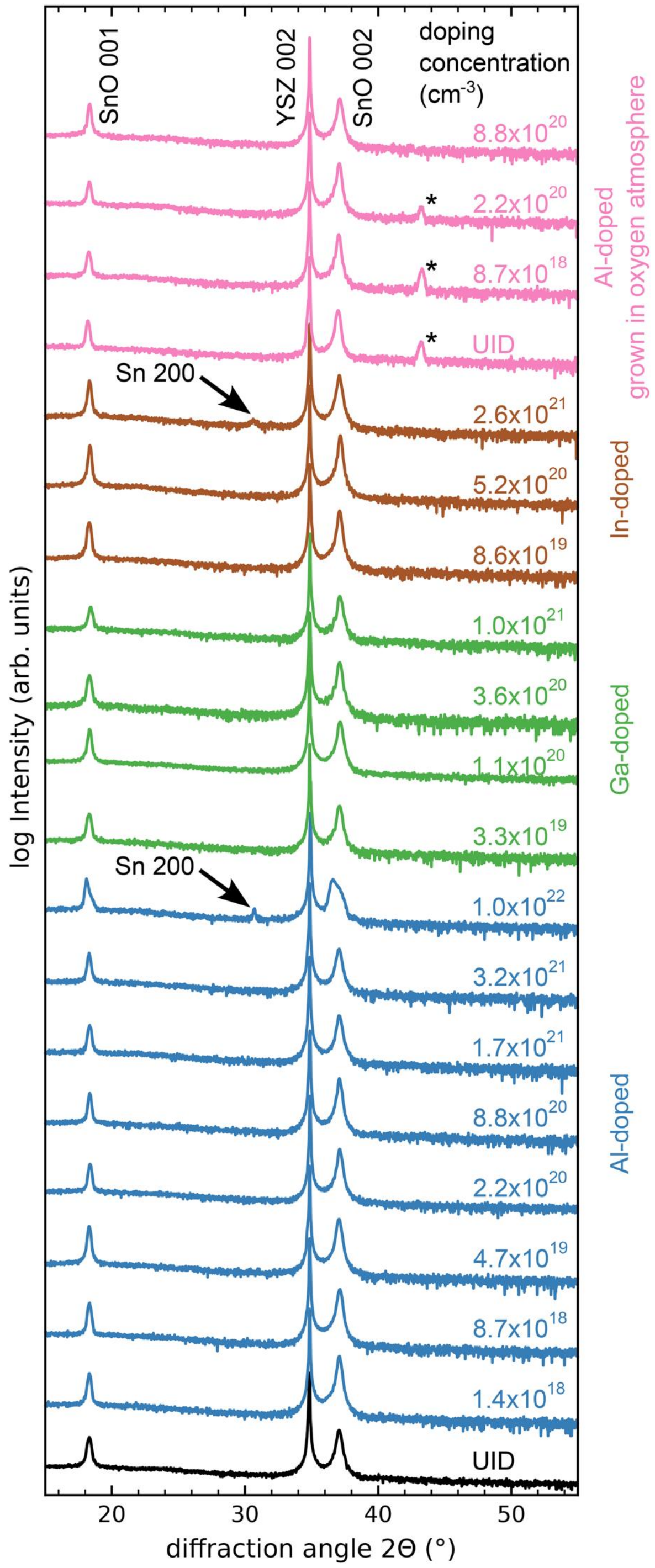

**FIG. 3.** X-ray diffraction (XRD) symmetric out-of-plane $2\theta$–$\omega$ scans of an unintentionally doped SnO sample (black), the Al-doped (blue), Ga-doped (green), In-doped (brown), and Al-doped samples grown in the oxygen atmosphere (pink). The nominal doping concentration is noted on the right side of the graph for each sample. Some samples show the Al 200 reflection from the Al-sample holder marked with $*$. A remeasured sample not showing the Al 200 reflection can be found in Fig. S8 of the supplementary material.

See the supplementary material for RHEED images of a substrate and a sample, additional information on the calculation of the nominal doping, further discussion on the assumed Al-doping mechanism, a description of the EDX procedure, the hole mobility plotted vs the hole concentration, further XRD measurements, XRR measurements and their fits for sample thickness determination, and AFM measurements of a substrate and several samples, including their discussion.

The authors thank H.-P. Schönherr and C. Hermann for their MBE support, and A. Gilbert for critically reading the manuscript. Our research was funded by the Federal Ministry of Research, Technology, and Space and the Berlin Senate.

## AUTHOR DECLARATIONS

### Conflict of Interest

The authors have no conflicts to disclose.

### Author Contributions

**N. Gutmann:** Conceptualization (supporting); Data curation (supporting); Formal analysis (lead); Investigation (lead); Methodology (equal); Project administration (equal); Validation (equal); Visualization (lead); Writing – original draft (lead); Writing – review & editing (equal). **G. Hoffmann:** Data curation (supporting); Investigation (supporting); Project administration (supporting); Supervision (supporting); Validation (lead); Writing – review & editing (equal). **A. F. Campbell:** Formal analysis (supporting); Investigation (supporting); Methodology (supporting); Software (lead); Validation (supporting); Writing – original draft (supporting); Writing – review & editing (supporting). **J. Lähnemann:** Formal analysis (supporting); Methodology (supporting); Software (supporting); Validation (supporting); Writing – review & editing (supporting). **O. Bierwagen:** Conceptualization (lead); Data curation (lead); Formal analysis (supporting); Methodology (equal); Project administration (equal); Supervision (lead); Validation (equal); Writing – review & editing (equal).

## DATA AVAILABILITY

The data that support the findings of this study are available within the article and its supplementary material. Hall- and XRD-raw data were generated at the NOMAD large scale facility. Derived data supporting the findings of this study are available from the corresponding author upon reasonable request.